\documentstyle[preprint,aps,epsf]{revtex}
\begin{document}
\preprint{APCTP/98-15}
\draft
\title{The Entropy of the BTZ Black Hole and \\
AdS/CFT Correspondence}
\author{Taejin Lee\cite{tlee}}
\address{Asia Pacific Center for Theoretical Physics, 207-43
Cheongryangri-dong\\
Dongdaemun-gu, Seoul 130-012, Korea \\ and \\
Department of Physics, Kangwon National University, 
                       Chuncheon 200-701, Korea} 
\date{\today}

\maketitle
\begin{abstract}
We construct an action, which governs the dynamics of 
the Ba\~nados-Teitelboim-Zanelli (BTZ) black hole and perform
the canonical quantization. The quantum action is given by a 
$SL(2,R)$ Wess-Zumino-Witten model on the boundary coupled to the 
classical anti-de Sitter background, representing a massless BTZ black hole.  
The coupling, determined by a one-cocyle condition, is found to 
give dominant contribution to the central charge of 
Virasoro algebra. The entropy of the BTZ black hole is discussed
from the point view of the AdS/CFT correspondence and an 
explanation is given to the puzzle of black hole entropy
in the BTZ case. The BTZ black hole is a quantum object and
the BTZ black hole with finite mass should be
considered as a quantum excitation of the massless one.
\end{abstract}

\pacs{PACS number(s): 04.60.K, 04.70.Dy, 11.25.Hf}


\section{Introduction}

The statistical interpretation of Bekenstein-Hawking entropy \cite{bh}
for the black hole has been one of the most outstanding problems in 
quantum gravity, since it may reveal what are the dynamical 
degrees of freedom one should count for in quantum gravity.
Recent studies on the black holes in the string theory brought
us one step closer to the heart of the question.
In the string theory one can construct the D-brane configurations
corresponding to the extremal black holes in five \cite{five} and four 
dimensions \cite{four}. Given that the number of BPS states is a
topological invariant, one can obtain the black hole entropy by 
counting the microstates in the weak coupling regime, which are
the BPS bound states of D-brane configuration. 
This was one of the remarkable achievements in the recent development of
string theory. There also have
been some attempts \cite{nonext} to extend this microstate counting 
to the non-extremal cases. However, the evaluation of the
entropy based on counting the BPS states in the weak coupling regime 
does not provide a satisfactory answer to the question of what are 
the microstates of the black hole, i.e., the dynamical degrees of gravity
in the strong coupling regime. And it is not directly 
applicable to the non-extremal black hole.

More accurate understanding of the microstates of the black hole 
follows from the recent discovery of the relationship of the BTZ
black hole \cite{btz,carlip95} in (2+1) dimensions and those 
higher dimensional
black holes in string theory. The intimate relationship
becomes clear as we observe that the higher dimensional black holes
can be transformed into the BTZ black hole configuration \cite{udual} 
by some dual transformations of the M-theory and
the near horizon geometry of the higher dimensional black holes
\cite{mal97} is described by the BTZ black hole. 
The advantage of utilizing 
this equivalence is that we can get more direct statistical 
interpretation of the black hole and this approach is not limited
to the extremal cases, since for the case of BTZ black hole
it may be possible to count explicitly the number of the microstates 
of the black hole regardless of whether it is extremal or not.
Thus, recently the BTZ black hole has stepped into the limelight in the
string theory \cite{lime}.  

In a recent work \cite{str97} Strominger suggested that the entropy
of BTZ black hole can be obtained by counting the number of the 
microstates in the conformal theory induced on the boundary of spatial
infinity. His evaluation of the entropy of the BTZ black hole 
is based on the work of Brown and Henneaux \cite{br86} where the 
generators for the diffeomorphism at asymptotic region are found to 
form Virasoro algebra with central charge 
$c=\frac{3l}{2G}$ where $-1/l^2$ is the cosmological constant
and $G$ denotes the Newton constant. Then, identifying 
the mass and angular momentum of the black hole in terms of 
Virasoro generators
and employing the asymptotic growth \cite{cardy} of the number of states 
in the corresponding conformal field theory, he obtains the 
Bekenstein-Hawking entropy of the BTZ black hole.
The statistical derivation of the entropy for the BTZ black hole
also has been discussed by Ba\~nados, Brotz and Ortiz \cite{bana98}
in their construction of the canonical partition function for
the BTZ black hole. However, there seems some discrepancy between
these recent works \cite{str97,bana98} and the previous ones 
\cite{carlip}. In a recent paper \cite{carlip98} Carlip
summarized the open problems associated with the known approaches
to the entropy of the BTZ black hole.

The purpose of the present work is to clarify some of the issues
raised in the literature on the BTZ black hole and
to provide an improved quantum theory of the BTZ black hole.
In the present paper we will derive the conformal field theory
on the boundary of spatial infinity, which describes the dynamical
degrees of freedom for the black hole, in the framework
of the Chern-Simons theory defined on the space-time with boundary.
The strategy to be employed is the one-cocycle construction
of Faddeev and Shatashivili \cite{faddeev86}.
The dynamics of the BTZ black hole is described 
by the Chern-Simons gauge fields in the bulk and
two sets of $SL(2,R)$ chiral WZW models coupled to the Chern-Simons
gauge fields on the boundary. The coupling of the chiral
WZW boson fields to the gauge fields is dictated
by the gauge invariance. It will be pointed out 
that the coupling of the WZW model to the Chern-Simons gauge
fields on the boundary plays an important role, giving the most dominant
contribution to the central charge. The importance of the coupling term 
was overlooked in the previous studies but was introduced later by 
hand in the form of Wilson line to produce the correct 
entropy in the construction of the grand-canonical partition 
function \cite{bana98}. 

The main result of the present study is that
in order to get the correct entropy for the BTZ black hole one must 
take into account the contribution of the gauge fields as well as the
WZW chiral bosons. In fact the former dominates over the latter.
The BTZ black hole is a quantum object, i.e.,
the BTZ black hole with finite mass should be
considered as a quantum excitation of the massless black hole
and the Bekenstein-Hawking entropy counts the density of quantum
states with respect to the vacuum state, being the massless
black hole.


\section{BTZ Black Hole}

The BTZ black hole with mass $M$ and angular momentum $J$ 
is described in the (2+1) dimensional gravity \cite{3dgr} 
by the following metric \cite{btz}
\begin{eqnarray}
ds^2_{BTZ} &=& -N^2 dt^2+ N^{-2} dr^2 + r^2(N^\phi dt+ d\phi)^2,\nonumber \\
N^2(r) &=& -M +\frac{r^2}{l^2} + \frac{J^2}{4r^2}, \label{btz}\\
N^\phi(r) &=& -\frac{J}{2r^2}. \nonumber
\end{eqnarray}
$N(r)$ has two roots, $r_+ > r_-$, which set the inner and outer
horizons of the black hole respectively.
If one introduces $SL(2,R)\otimes SL(2,R) \simeq SO(2,2)$ 
Lie algebra valued gauged fields 
$\quad A = A^{a}J_a, \quad \bar A = \bar A^{a}\bar J_a$
which can be written in terms of
the dreibein and the spin connection as follows;
$A_\mu{}^a = \omega_\mu{}^a + \frac{e_\mu{}^a}{l}, \quad
{\bar A}_\mu{}^a = \omega_\mu{}^a - \frac{e_\mu{}^a}{l}$,
we can describe the gravity in the presence of a cosmological
constant by a Chern-Simons action \cite{witten88}
\begin{eqnarray}
I_{CS}(A,\bar A) &=& \frac{k}{4\pi} \int_M \, {\rm tr} \, \left(AdA +
\frac{2}{3}AAA \right) 
- \frac{k}{4\pi} \int_M \, 
{\rm tr} \, \left(\bar{A}d\bar{A} + 
\frac{2}{3}\bar{A}\bar{A}\bar{A} \right) \label{csg}\\
\left[ J_a , J_b \right] &=& {\epsilon_{ab}}^c J_c, \quad
\left[ \bar J_a , \bar J_b\right] = {\epsilon_{ab}}^c \bar J_c, \quad
\left[ J_a , \bar J_b \right] = 0 \nonumber
\end{eqnarray}
where $k= -\frac{l}{4G} < 0$.
The equations of motion obtained from the Chern-Simons action
are satisfied by the BTZ black hole solution.

The BTZ black hole solution has two boundaries:
the horizon at $r=r_+$ and the boundary of spatial infinity.
Here we will mainly concern the boundary of spatial infinity.
If we confine ourselves to the region $r \ge r_+$, it is 
convenient to introduce a coordinates $(\tau, \rho, \phi)$ 
where $\tau = t/l$ and $r^2 = r^2_+ \cosh^2\rho - r^2_-
\sinh^2\rho$. In these new coordinate system the BTZ solution
reads as follows
\begin{mathletters}
\label{btzsol:all}
\begin{eqnarray}
&&\left ( \begin{array}{l}
 A_R = 2 \Delta R
 \left(\sinh\rho J_0 +\cosh\rho J_2\right),  \\
 A_L = 0, \\
 A_\rho = J_1, \\
\end{array} \right. \\
&&\left ( \begin{array}{l}
 \bar A_R = 0 \\
 \bar A_L = - 2 R
 \left(\sinh\rho \bar J_0 -\cosh\rho \bar J_2\right), \\
 \bar A_\rho = - \bar J_1. 
\end{array} \right.
\end{eqnarray}
\end{mathletters}
where 
\begin{eqnarray}
A_{R/L} &=& A_\tau \pm A_\phi, \quad
\bar A_{R/L} = \bar A_\tau \pm \bar A_\phi, \nonumber\\
\Delta R &=& \frac{(r_+-r_-)}{l},\quad R=\frac{(r_++r_-)}{l}.\nonumber
\end{eqnarray}

\section{Canonical Quantization}

In the presence of a boundary, we should impose appropriate boundary 
conditions and supplement some boundary terms.
We may choose the following boundary conditions, compatible with
the BTZ black hole solution, $A_L = 0$, $\bar A_R = 0$
and accordingly, the boundary 
terms, which do not pose a unitarity problem
\begin{eqnarray}
I_B = -\frac{k}{4\pi} \int_{\partial M}\, 
{\rm tr} (A_\tau - A_\phi) A_\phi
+\frac{k}{4\pi} \int_{\partial M}\, 
{\rm tr} (\bar A_\tau + \bar A_\phi) \bar A_\phi \label{bdaction}
\end{eqnarray}
where $\partial M$ denotes the boundary of spatial infinity.
As is well known, the gauge invariance is broken in the Chern-Simons
theory if the space-time has boundaries.
As a consequence, the degrees of freedom of the gauge fields 
corresponding to the broken symmetry become dynamical. 
The action for these "would be" gauge degrees of freedom
has been constructed as a one-cocycle in \cite{tlee97} by adopting the 
Faddeev-Shatashvili procedure \cite{faddeev86}:
\begin{eqnarray}
\alpha_{G}[A,\bar A,g,\bar g] 
 &=& I_G(A^g,\bar A^{\bar g})+I_B(A^g,\bar A^{\bar g}) 
 - I_G(A,\bar A)-I_B(A,\bar A)\\
 A^g &=& g^{-1} d g + g^{-1} A g \nonumber \\
 \bar A^{\bar g} &=& {\bar g}^{-1} d {\bar g} + 
 {\bar g}^{-1} {\bar A} {\bar g}. \nonumber
\end{eqnarray}
The explicit expression for the one-cocycle is given as
\begin{eqnarray}
\alpha_{G}(A, \bar{A},g,{\bar g})
&=& \alpha_1(A,g)+ {\bar \alpha}_1(\bar{A},{\bar g}), \label{acsg} \\
\alpha_1(A,g) &=& -\Gamma^L[g] - 
\frac{k}{2\pi}\int_{\partial M} {\rm tr}
(\partial_\phi g g^{-1})A_{L}, \nonumber\\
\bar \alpha_1(\bar A,\bar g) &=&  \Gamma^R[\bar g] + 
\frac{k}{2\pi}\int_{\partial M}{\rm tr} (\partial_\phi 
{\bar g}{\bar g}^{-1}){\bar A}_{R}, \nonumber \\
\Gamma^L[g] &=& \frac{k}{4\pi} \int_{\partial M} {\rm tr}
(g^{-1}\partial_- g)(g^{-1} \partial_\phi g) +\frac{k}{12\pi}
\int_{M} {\rm tr} (g^{-1} dg)^3, \nonumber\\
\Gamma^R[{\bar g}] &=& \frac{k}{4\pi} \int_{\partial M} {\rm tr}
({\bar g}^{-1}\partial_+ {\bar g})({\bar g}^{-1} \partial_\phi {\bar g}) 
+ \frac{k}{12\pi}\int_{M} {\rm tr} ({\bar g}^{-1} 
d{\bar g})^3 \nonumber
\end{eqnarray}
where $\partial_{\pm} = \partial_\tau \pm \partial_\phi$. 
(If $\partial M$ is the boundary of the horizon, one would replace $k$ 
by $-k$ in the action except for the coefficients of the WZ terms.) 
The induced action on $\partial M$
is given by a direct sum of two chiral WZW actions;
one with left moving chiral boson fields only
and the other with right moving chiral boson fields only.

Now the gauge symmetry is fully restored thanks to the
one-cocycle condition satisfied by $\alpha_{G}[A,\bar A,g,\bar g]$
\begin{eqnarray}
\delta \alpha_{G} &=& \alpha_{G}[A^h, \bar A^{\bar h}, g,\bar g] - 
\alpha_{G}[A,\bar A, hg, {\bar h}{\bar g}]
+\alpha_{G}[A, \bar A, h,\bar h]=0. \label{cocycle}
\end{eqnarray} 
Once a gauge condition and boundary conditions 
are chosen appropriately, the classical BTZ black hole solution
would be determined uniquely by the equations of motion. 
Thus, we may replace the gauge fields in the action by 
the classical BTZ solution in the classical limit. 
Then the second terms in $\alpha_1(A,g)$ and 
$\bar \alpha_1(\bar A,\bar g)$ depict 
couplings of the chiral boson fields to the classical BTZ black 
hole background.

The complete action for the BTZ black hole system consists of
the Chern-Simons terms, the boundary terms and the one-cocycle
\begin{eqnarray}
I_{G} (A, \bar{A},g,{\bar g}) &=& I_{CS} (A, \bar{A}) +
I_B (A, \bar{A}) + \alpha_{G}(A, \bar{A},g,{\bar g}) \\
&=& I_G[A,g]+ \bar{I}_G [\bar{A},\bar{g}]. \nonumber
\end{eqnarray}
It would be instructive to rewrite the action in the canonical form
\begin{eqnarray}
I_{G} &=&
\frac{k}{4\pi} \int_M {\rm tr} \epsilon^{ij}(\dot{A}_i A_j + A_\tau F_{ij})
-\frac{k}{4\pi} \int_{\partial M} {\rm tr} (g^{-1} \dot{g})
(g^{-1} g^\prime) - \frac{k}{12\pi} \int_M {\rm tr}
(g^{-1} dg)^3 \nonumber \\ 
& &  + \frac{k}{4\pi}\int_{\partial M} {\rm tr} \left\{
(A_\phi + g^\prime g^{-1})^2 
-2A_\tau (A_\phi + g^\prime g^{-1})\right\}, \label{can} \\
\bar{I}_{G} &=& -\frac{k}{4\pi} \int_M {\rm tr} 
\epsilon^{ij}(\dot{\bar A}_i {\bar A}_j + {\bar A}_\tau{\bar F}_{ij})
+\frac{k}{4\pi} \int_{\partial M} {\rm tr} ({\bar g}^{-1} 
\dot{{\bar g}}) ({\bar g}^{-1} {\bar g}^\prime) +
\frac{k}{12\pi} \int_M {\rm tr}({\bar g}^{-1} d{\bar g})^3 \nonumber\\ 
& & + \frac{k}{4\pi}\int_{\partial M} {\rm tr} \left\{
({\bar A}_\phi + {\bar g}^\prime {\bar g}^{-1})^2 
+2{\bar A}_\tau ({\bar A}_\phi + {\bar g}^\prime
{\bar g}^{-1})\right\} \nonumber
\end{eqnarray}
where ``$\cdot$" and ``$\prime$" denote derivatives with respect to
$\tau$ and $\phi$ respectively.

The fundamental Poisson brackets can be read as follows
\begin{eqnarray}
\{A^a_i({\bf x}), A^b_j({\bf x}^\prime)\} &=& -\frac{4\pi}{k}\epsilon_{ij}
\delta({\bf x}-{\bf x}^\prime)\eta^{ab} \label{poisson}\\
\{{\bar A}^a_i({\bf x}), {\bar A}^b_j({\bf x}^\prime)\} &=& 
\frac{4\pi}{k}\epsilon_{ij}
\delta({\bf x}-{\bf x}^\prime)\eta^{ab} \nonumber \\
\{A(\phi),B(\phi^\prime)\} &=& \frac{2\pi}{k} \delta(\phi-\phi^\prime)\,
{\rm tr}\, \left([a,b] g^\prime g^{-1}\right) +\frac{2\pi}{k} \delta^\prime
(\phi-\phi^\prime)\,{\rm tr}\,(ab) \nonumber\\
\{\bar{A}(\phi),\bar{B}(\phi^\prime)\} &=& -\frac{2\pi}{k} 
\delta(\phi-\phi^\prime)\,{\rm tr}\,\left(
[\bar{a},\bar{b}] {\bar g}^\prime \bar{g}^{-1}\right) 
-\frac{2\pi}{k} \delta^\prime (\phi-\phi^\prime)\,{\rm tr}\,(\bar{a}\bar{b}) 
\nonumber
\end{eqnarray}
where 
\begin{eqnarray}
A &=& {\rm tr\,} \left(a g^\prime g^{-1}(\phi)\right), \quad
B = {\rm tr}\, \left(b g^\prime g^{-1}(\phi)\right), \nonumber \\
\bar{A} &=& {\rm tr}\, \left(\bar{a} \bar{g}^\prime 
\bar{g}^{-1}(\phi)\right), \quad
\bar{B} = {\rm tr}\,\left(\bar{b} \bar{g}^\prime 
\bar{g}^{-1}(\phi)\right) \nonumber.
\end{eqnarray}
The Guass' constraints are given as
\begin{eqnarray}
\Phi &=&\frac{1}{2}\epsilon^{ij} F_{ij} 
-(A_\phi + g^\prime g^{-1})\delta\left({\rho-\rho_\infty}\right) = 0, \\
\bar{\Phi} &=&\frac{1}{2}\epsilon^{ij} {\bar F}_{ij} 
-({\bar A}_\phi + {\bar g}^\prime {\bar g}^{-1})
\delta\left({\rho-\rho_\infty}\right) = 0 \nonumber
\end{eqnarray}
where  $\rho_\infty \gg 1$ is the radius of the boundary of spatial 
infinity. The Hamiltonian is written as
\begin{eqnarray}
H =  -\frac{k}{4\pi}\int_{\partial M} {\rm tr} \left\{
(A_\phi + g^\prime g^{-1})^2+({\bar A}_\phi+{\bar g}^\prime 
{\bar g}^{-1})^2 \right\} 
\end{eqnarray}
and it satisfies $[H,\Phi^a] = 
[H,{\bar\Phi}^a]=0$.

We find that the generators for the local gauge symmetry 
can be written as
\begin{eqnarray}
G(\eta) &=& \frac{k}{4\pi} \int_M \eta^a \Phi_a =
\int_M \eta^a G_a - Q(\eta),\\
Q(\eta) &=& \frac{k}{2\pi} \int_{\partial M} {\rm tr}\, \eta 
\left(A_\phi+ g^\prime g^{-1}\right) \nonumber
\end{eqnarray}
where $G^a = \frac{k}{8\pi} \epsilon_{ij} F^a_{ij}$.
The surface term $Q$ has been introduced in the previous study
\cite{br86,bana98,bana94} to make $G$ differentiable.
Here we note that it arises as a consequence of canonical quantization
of the bulk-boundary system.
The conserved charges $Q$ generate the global (asymptotic) symmetries
of the theory. With the Poisson brackets
Eq.(\ref{poisson}), we find that the algebra of these global charges
have central extensions
\begin{mathletters}
\label{global1:all}
\begin{eqnarray}
\{G(\eta), G(\lambda)\} &=& G([\eta,\lambda])
+\frac{k}{4\pi} \int_{\partial M} \eta^a \left(
\lambda^\prime_a +\epsilon_{abc}A^b\lambda^c\right),\\
\{Q(\eta), Q(\lambda)\} &=& Q([\eta,\lambda]) + \frac{k}{4\pi} 
\int_{\partial M} \eta^a \left(\lambda^\prime_a
+\epsilon_{abc} A^b \lambda^c\right)
\end{eqnarray}
\end{mathletters}
where $([\eta,\lambda])^a = \epsilon^a{}_{bc}\eta^b \lambda^c$.

Similarly, we find the algebra of the global charges for the
right moving sector
\begin{eqnarray}
{\bar G}(\eta) &=& -\frac{k}{4\pi} \int_M \eta^a {\bar \Phi}_a =
\int_M \eta^a {\bar G}_a - {\bar Q}(\eta),\\
{\bar Q}(\eta) &=& -\frac{k}{2\pi} \int_{\partial M} {\rm tr}\, \eta 
\left({\bar A}_\phi+ {\bar g}^\prime {\bar g}^{-1}\right) \nonumber
\end{eqnarray}
where ${\bar G}^a = -\frac{k}{8\pi} \epsilon_{ij} {\bar F}^a_{ij}$.
They satisfy the following algebra
\begin{mathletters}
\label{global2:all}
\begin{eqnarray}
\{{\bar G}(\eta), {\bar G}(\lambda)\} &=& {\bar G}([\eta,\lambda])
-\frac{k}{4\pi} \int_{\partial M} \eta^a \left(
\lambda^\prime_a +\epsilon_{abc}{\bar A}^b\lambda^c\right),\\
\{{\bar Q}(\eta), {\bar Q}(\lambda)\} &=& {\bar Q}([\eta,\lambda]) - 
\frac{k}{4\pi} \int_{\partial M} \eta^a \left(\lambda^\prime_a
+\epsilon_{abc} {\bar A}^b \lambda^c\right).
\end{eqnarray}
\end{mathletters}

\section{Diffeomorphism on the boundary}

As discussed in ref.\cite{witten88} the diffeomorphism 
is equivalent to a gauge transformation with an appropriately 
chosen gauge function in the Chern-Simons formulation.
To be explicit the diffeomorphism generated by a vector
\begin{eqnarray}
\delta A_\mu{}^a = -\epsilon^\nu (\partial_\nu A_\mu{}^a 
-\partial_\mu A_\nu{}^a) -\partial_\mu(\epsilon^\nu A_\nu{}^a)
\end{eqnarray} 
is equivalent to a gauge transformation with a gauge function 
$\eta^a = -\frac{k}{4\pi} \epsilon^\mu A_\mu{}^a$ on shell.
The diffeomorphism on the boundary appears to be the asymptotic
symmetry and the corresponding global charges are identified
by the Virasoro generators. In order to identify the conformal
algebra let us consider a diffeomorphism generated by a
$\tau$-independent vector $\epsilon^\mu(\phi)$.
Under this surface diffeomorphism the gauge fields of the 
BTZ black hole solution transform as
\begin{eqnarray}
\delta A_\tau{}^0 &=& - \Delta R 
\epsilon_1 \cosh\rho,\nonumber \\
\delta A_\tau{}^2 &=& -\Delta R 
\epsilon_1 \sinh\rho,\nonumber\\
\delta A_\phi{}^0 &=& -\Delta R \left(
\epsilon_1 \cosh\rho+ \epsilon_+{}^\prime \sinh\rho \right),\label{tran1}\\
\delta A_\phi{}^1 &=& -\epsilon_1{}^\prime,\nonumber \\
\delta A_\phi{}^2 &=& -\Delta R \left(
\epsilon_1 \sinh\rho+ \epsilon_+{}^\prime 
\cosh\rho \right),\nonumber\\
\delta A_\tau{}^1 &=& \delta A_\rho{}^0 = 
\delta A_\rho{}^1 =  \delta A_\rho{}^2=0 \nonumber
\end{eqnarray}
where $\epsilon_+ = \epsilon_0+\epsilon_2$.
%

The diffeomorphism on the boundary of spatial infinity
is supposed to be the asymptotic symmetry. So it should be required
that $\delta A_i{}^a = 0$ at the leading order. From the asymptotic
behavior of $\delta A_\phi{}^a$,
\begin{eqnarray}
\delta A_\phi{} = -\frac{(J^0+J^2)}{2} \Delta R e^\rho \left[\epsilon_1
    + (\epsilon_+)^\prime\right] + {\cal O}(e^{-\rho})
\end{eqnarray}
it follows that
\begin{equation}
\epsilon_1 + (\epsilon_+)^\prime =0 \label{cond}.
\end{equation}
Note that it is not necessary to require $\delta A_\tau{}^a=0$ 
asymptotically, since $A_\tau{}^a$ are Lagrangian multipliers.

In order to transform the system by the two-dimensional diffeomorphism, 
we need to transform both the gauge fields and the chiral
boson fields: The gauge fields transform according to
Eq.(\ref{tran1}) with the condition Eq.(\ref{cond}) and the chiral
boson fields according to the usual diffeomorphism generated by
\begin{equation}
\delta \tau = \epsilon_0,\quad \delta\phi = \epsilon_2. \label{tran2}
\end{equation}   
If we transform the gauge fields 
\begin{eqnarray}
\delta_1 I_G &=& -\frac{k}{2\pi} \int_{\partial M} {\rm tr} \,
\epsilon^\prime_+\left\{ A_\phi(g^\prime g^{-1}) + J^1 
(g^\prime g^{-1})^\prime \right\}. 
\end{eqnarray}
And if we transform the boson fields,
\begin{eqnarray}
\delta_2 I_G &=& - \delta \Gamma^L [g] 
= -\frac{k}{2\pi} \int_{\partial M} {\rm tr} \, 
\epsilon^\prime_+ (g^\prime g^{-1})^2, 
\end{eqnarray}
Thus, we may identify the generators for the two-dimensional
diffeomorphism as follows
\begin{eqnarray}
\delta I_G(g) &=& - \frac{k}{2\pi}\int_{\partial M} 
\epsilon_+{}^\prime {\rm tr} \left\{(g^\prime g^{-1})^2 
+A_\phi (g^\prime g^{-1}) +(g^\prime g^{-1})^\prime J^1
\right\} \label{diff1}\\
\sum^\infty_{-\infty} L_n e^{in\phi} &=& -\frac{k}{2}{\rm tr}
\left\{(g^\prime g^{-1})^2 +A_\phi (g^\prime g^{-1}) +(g^\prime g^{-1})^\prime J^1 \right\}. \nonumber
\end{eqnarray}

Applying the same procedure we also find the variation of the
right moving part of the action.
Under the surface diffeomorphism, the gauge fields ${\bar A}$
transform as
\begin{eqnarray}
\delta{\bar A}_\tau{}^0 &=& R 
{\bar \epsilon}_1 \cosh\rho,\nonumber \\
\delta{\bar A}_\tau{}^2 &=& -R 
{\bar \epsilon}_1 \sinh\rho,\nonumber\\
\delta{\bar A}_\phi{}^0 &=& R \left(
-{\bar \epsilon}_1 \cosh\rho+ {\bar \epsilon}_-{}^\prime \sinh\rho \right),\\
\delta{\bar A}_\phi{}^1 &=& {\bar \epsilon}_1{}^\prime,\nonumber \\
\delta{\bar A}_\phi{}^2 &=& R \left(
{\bar \epsilon}_1 \sinh\rho- {\bar \epsilon}_-{}^\prime 
\cosh\rho \right),\nonumber\\
\delta{\bar A}_\tau{}^1 &=& \delta{\bar A}_\rho{}^0 = 
\delta{\bar A}_\rho{}^1 =  \delta{\bar A}_\rho{}^2=0 \nonumber
\end{eqnarray}
where ${\bar \epsilon}_- = {\bar \epsilon}_0-{\bar \epsilon}_2$.

From the leading behavior of $\delta {\bar A}_i$ in the asymptotic
region,
\begin{eqnarray}
\delta {\bar A}_\phi{} = -\frac{(J^0-J^2)}{2}
    R e^\rho \left[{\bar \epsilon}_1
    - ({\bar \epsilon}_-)^\prime\right] + {\cal O}(e^{-\rho}),
\end{eqnarray}
we obtain the condition for the surface diffeomorphism,
\begin{equation}
{\bar \epsilon}_1 - ({\bar \epsilon}_-)^\prime =0.
\end{equation}
Under this surface diffeomorphism the gauge fields ${\bar A}$
contributes to the variation of the action through the coupling
term as follows,
\begin{eqnarray}
\delta_1 I_G &=& \frac{k}{2\pi} \int_{\partial M} {\rm tr} \,
{\bar \epsilon}^\prime_-\left\{ {\bar A}_\phi({\bar g}^\prime {\bar g}^{-1}) 
- {\bar J}^1 ({\bar g}^\prime {\bar g}^{-1})^\prime \right\}.
\end{eqnarray}
As we transform the chiral boson fields ${\bar g}$
by the usual surface diffeomorphism generated by Eq.(\ref{tran2})
\begin{eqnarray}
\delta_2 I_G &=& \delta \Gamma^R [{\bar g}] 
= \frac{k}{2\pi} \int_{\partial M} {\rm tr} \, 
{\bar \epsilon}^\prime_- ({\bar g}^\prime {\bar g}^{-1})^2.
\end{eqnarray}
Taking into account transformation of both gauge fields
${\bar A}$ and chiral boson fields ${\bar g}$, we have
\begin{eqnarray}
\delta {\bar I}_G({\bar g})=
\frac{k}{2\pi}\int_{\partial M} {\bar\epsilon}_ -{}^\prime {\rm tr}\left\{
({\bar g}^\prime {\bar g}^{-1})^2+{\bar A}_\phi({\bar g}^\prime {\bar g}^{-1})- {\bar J}^1 ({\bar g}^\prime {\bar g}^{-1})^\prime \right\}.
\end{eqnarray}
Rewriting it in terms of Fourier modes, we get
the generators of the two-dimensional diffeomorphism  for the right moving chiral bosons
\begin{eqnarray}
\sum^\infty_{-\infty} {\bar L}_n e^{in\phi} &=& 
-\frac{k}{2} {\rm tr} \left\{
({\bar g}^\prime {\bar g}^{-1})^2+{\bar A}_\phi({\bar g}^\prime {\bar g}^{-1})- {\bar J}^1 ({\bar g}^\prime {\bar g}^{-1})^\prime \right\}
\label{diff2}.
\end{eqnarray}

\section{Entropy of the BTZ Black Hole}

According to the Bekenstein-Hawking formula the thermodynamic entropy of a
black hole is proportional to the area of the event horizon \cite{bh}:
$S = \frac{A}{4\hbar G}$ where $A$ is the area of the horizon and $G$ is
the Newton constant. The Bekenstein-Hawking entropy suggests that it
has a statistical interpretation. It also implies that the statistical
interpretation must be of classical origin, since the quantum corrections
contribute to the entropy only at the order of 1. On the other hand, it 
contradicts the no-hair theorem, which asserts that the classical 
state of the black hole can be uniquely determined by its 
conserved charges measured asymptotically.
Since the seminal works of Bekenstein and Hawking 
the statistical interpretation of the black hole entropy
has been one of the outstanding problems in theoretical physics. 
It has been believed long that understanding the black hole 
entropy and related problems may provide clue to the consistent 
quantum theory of gravity. 
Since the Chern-Simons gravity is regarded as a consistent 
quantum theory for gravity in (2+1) dimensions, 
it is supposed to provide a solution to
this conundrum. Here in this section we will discuss how the
puzzle can be resolved in the framework of the Chern-Simons gravity.

In the semi-classical region, where the cosmological constant
is small, $k \gg 1$ the entropy of the black hole must be 
given by the Bekenstein-Hawking formula 
$S = \frac{A}{4\hbar G}$ where $A$ is the area of the horizon.
For the BTZ black hole it reads as 
\begin{equation}
S=\frac{2\pi r_+}{4G}= \sqrt{2}\pi\left(M+\sqrt{M^2-(J/l)^2}
\right)^{1/2}.
\end{equation} 
Since the BTZ black hole is described by a $SL(2,R)$ WZW model
as discussed in the previous sections,
it is suggestive that the entropy may be obtained by counting
the microstates in the conformal field theory on the boundary.
Given $n_L$ and $n_R$, the eigenvalues of the Virasoro generators 
$L_0$ and ${\bar L}_0$,  
the asymptotic growth of the states is given by the Cardy's
formula \cite{cardy} when $n_L \gg c_L$ and $n_R \gg c_R$. 
Then, the entropy may be given as
\begin{equation}
S= 2\pi \sqrt{\frac{c n_L}{6}} + 2\pi \sqrt{\frac{c n_R}{6}}.
\end{equation}

In ref.\cite{str97} Strominger derived 
the entropy of the BTZ black hole, employing the Cardy's
formula and the previous study on (2+1) dimensional AdS space
by Brown and Henneaux. 
Adopting the analysis of ref.\cite{br86} on (2+1) dimensional 
gravity where the generators for the diffeomorphism at 
asymptotic region are found to 
form a Virasoro algebra with central charge 
$c=\frac{3l}{2G}= -6k$, identifying the mass and angular 
momentum of the black hole as follows
\begin{eqnarray}
M = \frac{8G}{l} (L_0 + {\bar L}_0), \quad J = L_0 - {\bar L}_0, \label{mass}
\end{eqnarray}
he found that the Cardy's formula yields the
Bekenstein-Hawking entropy of the BTZ black hole exactly.

Given a conformal field theory for the BTZ black hole at hand,
it looks straightforward to give a concrete form 
to the Strominger's derivation. 
However, one may readily realize that the generators for the two-dimensional
diffeomorphism in Eqs.(\ref{diff1},\ref{diff2}) do not form the Virasoro
algebra contrary to expectations except when $A_\phi=0$ on $\partial M$.
Moreover, the expectation values of the Virasoro generators 
$L_0$ and ${\bar L}_0$ do not satisfy the Eq.(\ref{mass}).
However, if we discern that the system faithfully respects the
AdS/CFT (anti-de Sitter space/conformal field theory) correspondence 
\cite{mal97,ads}, we can find a resolution.

Note that since the Guass' law $F={\bar F}=0$ is imposed for the gauge field
in the bulk, we may write the gauge fields 
as $A = (u_0u)^{-1} d(u_0u)$, 
${\bar A} =({\bar u}_0 {\bar u})^{-1} d({\bar u}_0{\bar u})$ with some
multi-valued gauge functions $u$ and ${\bar u}$ \cite{cangemi92}.
Here the gauge field configuration, $ \{A_0=u^{-1}_0 du_0,\,\,\,
{\bar A}_0={\bar u}^{-1}_0 d{\bar u}_0 \} $, depicts the BTZ black hole
with zero mass.
If we substitute it for the gauge fields in the action Eq.(\ref{can}),
\begin{mathletters}
\label{action:all}
\begin{eqnarray}
I_G [A,g] &=& -\Gamma^L [u_0u] -\Gamma^L[g] -\frac{k}{2\pi} \int_{\partial M}
{\rm tr} (g^\prime g^{-1}) (u^{-1}\partial_- u), \label{action:a} \\
{\bar I}_G [{\bar A},{\bar g}] &=& \Gamma^R [{\bar u}_0{\bar u}] +
\Gamma^R[{\bar g}]+\frac{k}{2\pi} \int_{\partial M}
{\rm tr} ({\bar g}^\prime {\bar g}^{-1}) ({\bar u}^{-1}\partial_+ {\bar u}).
\end{eqnarray}
\end{mathletters}
Making use of the Polyakov-Wiegman identity,
\begin{eqnarray}
\Gamma^{R/L}[ug] = \Gamma^{R/L}[u] +\Gamma^{R/L}[g]
+ \frac{k}{2\pi} \int_{\partial M} {\rm tr} 
(g^\prime g^{-1}) (u^{-1}\partial_\pm u),
\end{eqnarray}
or the one-cocycle condition Eq.(\ref{cocycle}), we find
\begin{eqnarray}
I_G [A,g] = -\Gamma^L [u_0]-\Gamma^L [ug], \quad
{\bar I}_G [{\bar A},{\bar g}] = 
\Gamma^R [{\bar u}_0]+ \Gamma^R [{\bar u}{\bar g}]. \label{action2}
\end{eqnarray}
This equation implies that the system can be equivalently described
by the boundary fields $h= ug$ and ${\bar h}= {\bar u}{\bar g}$,
which coupled to a trivial background, i.e.,
the massless BTZ black hole, but may have non-trivial holonomies in 
contrast to $g$ and ${\bar g}$. Also it reveals
that the generating functional defined as
\begin{eqnarray}
Z[A|, {\bar A}|] = \int D[A] D[{\bar A}] D[g]D{\bar g}]
\exp i \left[I_G [A,g] +{\bar I}_G [{\bar A},{\bar g}] \right],
\end{eqnarray}
does not change under small variation of the boundary values of the 
gauge fields $A|$, ${\bar A}|$ and the bulk degrees 
of freedom perfectly match the boundary degrees of freedom.
This may be considered as the simplest realization of
the AdS/CFT correspondence \cite{tlee98}.

Let us return to the boundary conformal field theory. 
We may develop the quantum theory in terms of the $h$ and
${\bar h}$ in parallel with the previous sections;
as a result we may simply replace $g$ with $h$ and the 
gauge fields $A$ with the those of massless black hole solution.
Here the vacuum state corresponds to the massless BTZ black hole.
In the canonical quantization the global charges become
\begin{eqnarray}
Q(\eta) = \frac{k}{2\pi} \int_{\partial M} {\rm tr}\, \left(\eta 
h^\prime h^{-1}\right), \quad
{\bar Q}(\eta) = -\frac{k}{2\pi} \int_{\partial M} {\rm tr}\, \left(\eta 
{\bar h}^\prime {\bar h}^{-1}\right)
\end{eqnarray}
and satisfy the usual Kac-Moody algebra
\begin{mathletters}
\label{kmalg:all}
\begin{eqnarray}
\{Q(\eta), Q(\lambda)\} &=& Q([\eta,\lambda]) + \frac{k}{4\pi} 
\int_{\partial M} \eta^a \lambda^\prime_a \\
\{{\bar Q}(\eta), {\bar Q}(\lambda)\} &=& {\bar Q}([\eta,\lambda]) - 
\frac{k}{4\pi} \int_{\partial M} \eta^a \lambda^\prime_a.
\end{eqnarray}
\end{mathletters}
The algebra of the global charges can be obtained
also by applying the symplectic reduction method and the Noether 
procedures \cite{oh98}.
Expanding it in terms of Fourier modes, we have
\begin{mathletters}
\label{km:all}
\begin{eqnarray}
\{T^a_n, T^b_m\} &=& \epsilon^{ab}{}_c T^c_{n+m} + 
i\frac{k}{2} \eta^{ab} n \delta (n+m), \\
\{{\bar T}^a_n, {\bar T}^b_m\} &=& \epsilon^{ab}{}_c {\bar T}^c_{n+m} - 
i\frac{k}{2} \eta^{ab} n \delta (n+m), \\
T^a_n &=& \frac{k}{2\pi} \int_{\partial M} {\rm tr} J^a 
(h^\prime h^{-1}) e^{-in\phi} d\phi, \nonumber\\
{\bar T}^a_n &=& -\frac{k}{2\pi} \int_{\partial M} {\rm tr} {\bar J}^a 
({\bar h}^\prime {\bar h}^{-1}) e^{-in\phi} d\phi .\nonumber
\end{eqnarray}
\end{mathletters}

Now the action transforms under the two-dimensional 
diffeomorphism as
\begin{mathletters}
\begin{eqnarray}
\delta I_G(h) &=& - \frac{k}{2\pi}\int_{\partial M} 
\epsilon_+{}^\prime {\rm tr} \left\{(h^\prime h^{-1})^2 
+(h^\prime h^{-1})^\prime J^1 \right\}, \\
\delta {\bar I}_G({\bar h}) &=& \frac{k}{2\pi}\int_{\partial M} 
{\bar \epsilon}_-{}^\prime {\rm tr} \left\{({\bar h}^\prime 
{\bar h}^{-1})^2 -({\bar h}^\prime {\bar h}^{-1})^\prime {\bar J}^1 \right\}
\end{eqnarray}
\end{mathletters}
and the generators for the two-dimensional diffeomorphism read as
\begin{mathletters}
\label{dd:all}
\begin{eqnarray}
\sum^\infty_{-\infty} L_n e^{in\phi} &=& - \frac{k}{2}{\rm tr}
\left\{(h^\prime h^{-1})^2 +(h^\prime h^{-1})^\prime J^1 \right\} \\
\sum^\infty_{-\infty} {\bar L}_n e^{in\phi} &=& -\frac{k}{2}{\rm tr}
\left\{({\bar h}^\prime {\bar h}^{-1})^2 -
({\bar h}^\prime {\bar h}^{-1})^\prime {\bar J}^1 \right\}.
\end{eqnarray}
\end{mathletters}
Note that if the one-loop radiative corrections of the $SL(2,R)$ Chern-Simons 
theory \cite{witten89} is taken into consideration, 
$k$ in Eqs.(\ref{km:all},\ref{dd:all}) would be shifted by 2. 
(This radiative correction, however can be ignored in the large $k$ limit.)

Rewriting the diffeomorphism generators in terms of the $SL(2,R)$ currents,
$T^a_n$ and ${\bar T}^a_n$, 
\begin{mathletters}
\begin{eqnarray}
L_n &=& -\frac{1}{k+2} \sum_m T^a_{n-m} T^a_m - i \frac{n}{2} T^1_n, \\
{\bar L}_n &=& -\frac{1}{k+2} \sum_m {\bar T}^a_{n-m} {\bar T}^a_m 
- i \frac{n}{2} {\bar T}^1_n,
\end{eqnarray}
\end{mathletters}
we find that they form the Virasoro algebra
\begin{mathletters}
\begin{eqnarray}
[L_n, L_m] &=& (n-m)L_{n+m} + \frac{c_L}{12} n(n^2-1)\delta (n+m), \\
\left[ {\bar L}_n, {\bar L}_m \right] &=& (n-m){\bar L}_{n+m} + 
\frac{c_L}{12} n(n^2-1)\delta (n+m)
\end{eqnarray}
\end{mathletters}
with central charges, $c_L = c_R = \frac{3k}{k+2}-6k$.
Here the shift in $k$ may be considered as a consequence of the
the radiative correction or the Sugawara construction of the
energy-momentum tensor.
It is interesting to observe that the algebraic structure of the conformal
theory for the BTZ black hole is identical to that of the two-dimensional
gravity \cite{2dg}. In the large $k$ limit the central charges are
approximated by $-6k$.

The evaluation of the entropy would be completed by taking expectation
values of $L_0$ and ${\bar L}_0$. Observe that
\begin{eqnarray}
{\rm tr} (h^\prime h^{-1})^2 &=& {\rm tr} 
\left(u^{-1}u^\prime+g^\prime g^{-1}\right)^2, \nonumber\\
{\rm tr} ({\bar h}^\prime {\bar h}^{-1})^2 &=& {\rm tr}
\left({\bar u}^{-1}{\bar u}^\prime+{\bar g}^\prime 
{\bar g}^{-1}\right)^2. \nonumber
\end{eqnarray}
We take the expectation values of $L_0$ and ${\bar L}_0$ with respect
to the quantum state corresponding to the BTZ black hole with mass $M$ and 
angular momentum $J$. For the BTZ black hole state 
the expectation values of $(u^{-1}u^\prime)$ and 
$({\bar u}^{-1}{\bar u}^\prime)$ take $(A_{BTZ})_\phi$ 
and $({\bar A}_{BTZ})_\phi$ of the classical BTZ solution Eq.(\ref{btzsol:all}).
If neglecting the excitation in $g$, we have 
\begin{mathletters}
\begin{eqnarray}
n_L &=& \bigl<BTZ|L_0|BTZ \bigr> = 
-\frac{k}{2} {\rm tr} \left[(A_{BTZ})_\phi\right]^2 
= -\frac{k}{4}\left(M-\frac{J}{l}\right), \\
n_R &=& \bigl<BTZ|{\bar L}_0|BTZ \bigr> = 
-\frac{k}{2} {\rm tr} \left[({\bar A}_{BTZ})_\phi\right]^2 
=- \frac{k}{4}\left(M+\frac{J}{l}\right). 
\end{eqnarray}
\end{mathletters}
Thus, employing the Cardy's formula, 
we obtain the correct entropy of the
BTZ black hole
\begin{eqnarray}
S &=& \pi |k| \left(\sqrt{M-\frac{J}{l}} +\sqrt{M+\frac{J}{l}}\right)\\
&=& \frac{2\pi r_+}{4G}. \nonumber
\end{eqnarray}
(When we employ the Cardy's formula, we assume that the
conformal field theory is unitary. However, $SL(2,R)$ group
does not have a finite unitary representation. In order to
get finite unitary representations, one may gauge a $U(1)$
subgroup \cite{tlee97}. This resluts in $SL(2,R)/U(1)$
WZW model for the BTZ black hole, and shift of the 
central charges by 1. But this shift is negligible in
the large $k$ limit.)

The Virasoro algebra of the surface diffeomorphism
was first derived by analyzing the asymptotic symmetry group of the 
AdS space in the ADM formulation \cite{br86} and later
by constructing the global charges corresponding to the diffeomorphism in the
Chern-Simons formulation of the (2+1) dimensional gravity \cite{bana94}.
In the present paper we give a direct derivation, compatible with
the AdS/CFT correspondence, by examining the variation of the
action under the surface diffeomorphism.
The analysis given in the present 
paper points out that the surface diffeomorphism should be 
considered in the context of the three dimensional diffeomorphism. 
So when we take the transformation on the boundary, 
$\delta \tau =\epsilon_0$,
$\delta \phi = \epsilon_2$, we should also transform
the boundary values of the gauge fields consistently; 
otherwise the central charge of the Virasoro algebra would be
obtained as $\frac{3k}{k+2}$. Thus, the coupling of the bulk
field degrees of freedom to the boundary ones plays an important
role, and moreover, it gives dominant contribution to the
central charge in the semi-classical regime. The analysis
given in this section also clearly exhibit that one must
consider the gauge field degrees of freedom, which often  
ignored, as a part of the dynamical degrees of freedom of 
the black hole. In fact the entropy of the black hole
is mainly comprised of the contribution from the gauge field
degrees of freedom.

\section{Conclusions}

Understanding the Bekenstein-Hawking entropy of the black hole 
from the first principles has been the crux of
gravity, and is believed to guide us eventually to a consistent
quantum theory of gravity. Since in (2+1) dimensions the gravity
can be formulated by a Chern-Simons theory, which is supposed
to be finite, this open problem may be treated in a tractable
manner in (2+1) dimensions. Recent development of the string 
duality and the M-theory points out that the three 
dimensional BTZ black hole is
extremely important to study the higher dimensional black holes
in the string theory in that many of them contain the BTZ black hole
as a part of their near-horizon geometries. 
In the present paper we discuss the entropy of the BTZ black hole
in the Chern-Simons formulation and identify the relevant
dynamical degrees of freedom of the black hole.
The derivation of the BTZ black hole entropy given in this work
is consistent with the AdS/CFT correspondence and clarifies
the issues raised in the literature.

We summarize the present work and conclude with discussions on
the puzzle of the black hole entropy.
We construct an action, which governs
the dynamics of the BTZ black hole and perform the canonical
quantization. Then we identify the global charges,
which generate the asymptotic symmetry of the system,
in the framework of the canonical quantization and find that the
algebra of these charges have central extensions. 
The generators of the surface diffeomorphism are directly obtained
from the variation of the action under the diffeomorphism,
which comprises the transformation of the chiral boson fields and
the gauge fields on the boundary. The coupling between the
chiral boson fields and the gauge fields on the boundary
is found to play an important role, giving dominant contribution
to the central charge.

In the presence of the coupling term, we can apply the 
Polyakov-Wiegman identity to the action, and find that the
system can be equivalently described by chiral boson fields with
non-trivial holonomies, which coupled to the background given as
the massless black hole. In this new setting, the
global charges satisfy the Kac-Moody algebra with
central extension at the classical level and the 
diffeomorphism generators form a Virasoro algebra as desired.
From this point of view, the appropriate vacuum state for the
quantum theory is the massless black hole rather than the classical
BTZ black hole configuration with finite mass:
The state of massless black hole is the eigenstate of 
$L_0$ and ${\bar L}_0$ with zero eigenvalues. 
Accordingly, the BTZ black hole with finite mass should
be understood as a quantum excitation of the massless black hole.
And the Bekenstein-Hawking entropy counts the density of quantum
states at the BTZ state with respect to the vacuum state 
of the massless black hole.

The present derivation also provides an appropriate
explanation to the puzzle of the black hole entropy, which
briefly mentioned at the beginning of the section V.
Although the Bekenstein-Hawking entropy is proportional
to $1/\hbar$, it is of quantum mechanical origin.
In other words the mass and angular momentum of the
black hole are quantum hairs, which unlike the small 
fluctuation of $g$, have large macroscopic expectation
values. Hence, the classical no-hair theorem does not apply.
The relevant dynamical degrees of freedom,
which give main contribution to the entropy
are not the chiral boson 
fields $g$, which only describe small fluctuations, but
the gravity, here the Chern-Simons gauge fields:
The gauge fields of the classical solution should be 
interpreted as expectation values of the gauge fields 
in the quantum theory. This interpretation also gives a proper 
answer to the question of where the relevant degrees of freedom 
of the black hole are located. The true meaning of the 
Bekenstein-Hawking black hole entropy can be found
only when the gravity is fully quantized. 
We hope that this work also sheds some light on the study 
of the D-brane dynamics.

\acknowledgments

This work was supported in part by the Basic Science Research 
Institute Program, Ministry of Education of Korea (BSRI-97-2401)
and by KOSEF through CTP at SNU. The author would like to thank 
Professor R. Jackiw for introducing ref.\cite{oh98} to him and
S. Hyun and J.-H. Cho for useful discussions. Part of the work
was done during the author's visit to ICTP.

\end{document}